# An Information Network Overlay Architecture for the NSDL

Carl Lagoze
Cornell Information Science
301 College Ave.
Ithaca, NY 14850
+1-607-255-6046

lagoze@cs.cornell.edu

Dean B. Krafft
Cornell Computer Science
308 Upson Hall
Ithaca, NY 14853
+1-607-255-9214

dean@cs.cornell.edu

Susan Jesuroga
UCAR-NSDL
PO Box 3000
Boulder, CO 80307
+1-303-497-2942

jesuroga@ucar.edu

Tim Cornwell
Cornell Information Science
301 College Ave.
Ithaca, NY 14850
+1-607-255-3297

cornwell@cs.cornell.edu

Ellen J. Cramer
Cornell Information Science
301 College Ave.
Ithaca, NY 14850
+1-607-255-3297

elly@cs.cornell.edu

Edwin Shin
Cornell Information Science
301 College Ave.
Ithaca, NY 14850
+1-607-255-5587

eddie@cs.cornell.edu

## ABSTRACT
We describe the underlying data model and implementation of a new architecture for the National Science Digital Library (NSDL) by the Core Integration Team (CI). The architecture is based on the notion of an *information network overlay*. This network, implemented as a graph of digital objects in a Fedora repository, allows the representation of multiple information entities and their relationships. The architecture provides the framework for contextualization and reuse of resources, which we argue is essential for the utility of the NSDL as a tool for teaching and learning.

## Categories and Subject Descriptors
H.3.7 [**Information Storage and Retrieval**]: Digital Libraries – *collection, standards, system issues, user issues.*

## 1. INTRODUCTION
From the inception of the NSDL Project in 2000 through 2003, the mission of the Core Integration (CI) team of the NSDL focused on building the architecture and organization necessary to deploy the NSDL main portal. This portal provided basic digital library functionality including search, browse, and access to resources from both NSDL-funded collections and various external organizations. In addition, the underlying architecture supported other services including resource archiving, user authentication, and a virtual reference desk. The initial public launch of the main NSDL portal took place at the 2002 NSDL annual meeting. Follow-on work through 2003 concentrated on rationalizing and automating procedures underlying the portal. This first-phase work accomplished a number of goals including demonstrating a proof-of-concept, establishing a public presence, and providing motivation for other NSDL projects to put resources on-line and garner momentum for the future development of the NSDL.

The successful realization of these initial tasks set the stage for CI to refine its mission with sharper focus on the specific educational goals of the overall NSDL project. Broadly speaking, these goals include establishing the technical, community, and organizational frameworks for improving Science, Technology, Engineering, and Mathematics (STEM) education through innovative use of the Web and internet [39]. Implicit in these goals is the notion that the NSDL is not only a "digital library", but it is a library with special characteristics reflecting its focus on education.

One manifestation of this refined mission is NSF funding in late 2004 of a new set of projects, the Pathways projects [4], charged with defining and deploying the NSDL for specific learning communities. These projects build on the notion that the NSDL should have multiple "faces" to meet the divergent quality, user interface, resource, and support requirements of the broad NSDL target communities (essentially the entire breadth of STEM learning and teaching communities).

The creation of these separate pathways into the NSDL entails more than just defining different user interfaces or establishing different resource selection criteria. While these are important, it is arguably more important to augment NSDL resources with context that defines their usability and reusability in different learning and teaching environments. By "context", we mean information such as the manner in which resources have been used (e.g., in what curricula and lesson plans), logs showing who used them and how successful those uses were, comments by teachers and fellow learners that annotate and explain primary resources, and linkages between the resources and relevant state educational standards. Instructional technology experts like Tom Reeves [30] have argued that without such context, the educational value of a digital library is severely compromised.

To represent such contextual information, CI must define and support a more expressive and extensible architectural framework than the metadata-based framework implemented in the existing Metadata Repository (MR) [16]. This framework must represent a range of entities beyond metadata, such as content, agents (e.g., people and organizations), services, standards, subject areas, and the like. It must represent the web of relationships among these entities, which provide the contexts for effective resource use. Finally, it must accommodate contributions from multiple participants and allow fine-grained access to the information it represents.

This paper describes recent work by CI on a new technical infrastructure based on the notion of an *information network overlay architecture.* The deployment date for this work is the third quarter of 2005. This architecture represents the data underlying the NSDL as a graph of typed nodes, corresponding to the multiple entities comprising the NSDL, and semantic edges representing the contextual relationships among those entities. The information nodes in this network may be *internal,* containing information such as a Dublin Core record harvested from an NSDL metadata provider. Nodes may also be *redirect,* containing a URL that maps to information at some networked location. Finally, nodes may be *virtual*, representing information output from web services that process other information represented in the network. The result is an information environment for contextualizing, adding value, and dynamically reusing distributed resources.

The new NSDL information network is implemented within Fedora [17], an open-source content-management and repository system. Fedora has a number of features that allow the creation of this network, including the aggregation of local and distributed content, the association of web services with aggregated content to produce dynamic representations of the content, and the representation of relationships among those content objects. A Fedora repository is implemented as a set of web services, and consequently all its functionality is exposed through SOAP and REST APIs. It thereby provides full programmatic access to the data model and integrates into an extended web services infrastructure. As a result, portal builders such as the Pathways projects can easily derive and tailor information in the CI Fedora repository to create manifestations of the NSDL specific to their target audiences.

The remainder of this paper is organized as follows. Section 2 describes the relevance of contextualization and subsequent reuse for digital libraries serving education. Section 3 provides an overview of the initial NSDL architecture, with the goal of understanding its successes and limitations. Section 4 is the core of the paper, describing the design of the information network overlay and its implementation in Fedora. Section 5 provides conclusions and future directions.

## 2. CONTEXT AND REUSE

Digital libraries offer universal access to online resources and hold the promise of widespread reuse of digital resources for education. While accessing quality resources and employing good technology are important, digital libraries need to support the full life cycle of data, information, and knowledge, and knowledge construction in general[19]. Thinking of digital libraries merely as repositories is too limiting for effective use in education. Reeves wrote "The real power of media and technology to improve education may only be realized when students actively use them as cognitive tools rather than simply perceive and interact with them as tutors or repositories of information." [29]. Looking to research in the educational technology community provides important perspectives for understanding the challenges of successful resource reuse.

The common term to describe reusable resources in the educational community is "learning objects," whose notion is grounded in the object-oriented model of computer science [37]. When an educator or computer system finds and reuses a learning object, they ultimately place the object into a specific learning context. A learning context includes many dimensions including social and cultural contexts; the context of the learner's educational system; and the learner's abilities, preferences and prior knowledge [20]. For the NSDL, learning context presents a significant challenge because of the breadth of audiences served and the differences in the domains of knowledge represented.

Most digital libraries currently rely on forms of metadata to describe learning objects and enable discovery. Metadata standards abstract properties of learning objects, and abstraction can lead to instances where the context is ignored or overly simplified [23]. Taxonomies, vocabularies, and ontologies are being developed to improve consistency of metadata and search results. Yet the resulting metadata is focused on the technical aspects of description and cataloging, not on capturing the actual context of instructional use. Recker and Wiley write "a learning object is part of a complex web of social relations and values regarding learning and practice. We thus question whether such contextual and fluid notions can be represented and bundled up within one, unchanging metadata record." [28]

McCalla also argues that there is no way of guaranteeing that metadata captures the breadth and depth of content domains. He writes that, ideally, learning objects need to reflect "appropriateness" to address the differences between learners' needs. [21] In addition, questions remain as to whether these logical representations (e.g. metadata and vocabularies) created primarily for use by computer systems will make the most intuitive sense for learners [9]. Koper observes that using open standards, such as SCORM and IMS Content Packaging, "without human interpretation and the provision of a manual it is not possible to interpret the pedagogical structure." [15]

Within current educational technology research, several approaches have been suggested to help supply context. Parrish suggests capturing opinions about learning objects, descriptions of how they are used, and a history of use and users to enable learning object reuse [23]. Recker and Wiley recommend a system that includes multi-record, non-authoritative information describing the community of users from which the learning object is derived [28]. McCalla describes an ecological approach to capture learner interactions and attach models of learners to the learning objects with which they interact [21]. Allert outlines a modeling approach to describe "Learner Roles" and proposes that, with such information, a learning object can dynamically adopt properties from such roles to provide context [6].

In the digital library community, Recker, et al. found middle and high school teachers were interested in 1) teacher-recommended resources and teacher-created repositories, as opposed to generic digital resources; and 2) resources linked to state education standards [27]. Abbas, et al. suggests that tracking and using student-generated search keywords might enable systems to provide more age-appropriate representations of resources [5]. McMartin and Terada reported that NEEDS resource authors felt access to comments or reviews by other faculty and students would be the most useful digital library service [22].

Sumner also proposed new models for digital libraries, which could scaffold the sense-making activities of learners by helping them make use of library resources to construct their own knowledge representations [35]. In evaluating the GetSmart system, Marshall, et al. wrote that effective learning strategies can be supported by combining curriculum tools, concept mapping (node-link representations of knowledge) and search tools on top of digital libraries [19].

This short review of the literature suggests that the NSDL must enable contextualization of resources to ensure adoption and successful reuse. Future NSDL architectures must support not only resources and descriptive metadata, but also the relationships between these resources and a wide range of contextual information about learners, their needs, and the learning environment and community they are situated within.

## 3. STARTING WITH A LIMITED WORLD VIEW

A good model for the phase I metadata repository-based implementation of the NSDL is the notion of a *union catalog*. In this model, the NSDL consists of the union of sets of resources from individual collections. The collections themselves are carefully vetted and chosen to ensure that they are providing high quality educational resources. This, in turn, guarantees that the resources in the NSDL will be valuable and appropriate to STEM education.

Each of the selected collections contributes its catalog records (metadata) to the central NSDL site, including pointers to the digital resources described in these catalog records. The NSDL aggregates these records, ensures that they are in a standard, normal format, and provides a central location from which users can search or browse the union of all the catalog records from each of the collections. In addition, the NSDL CI serves as a central point of redistribution of these records. This greatly simplifies the provision of other services (e.g. resource archiving) because those services need only interact with a single, standard location and service to obtain the information that they need.

This initial model is a good first step toward meeting the primary goal of the NSDL: enabling easy discovery of appropriate STEM digital resources by students, teachers, researchers, and the general public. Through the nsdl.org portal, the NSDL provides a "one-stop" interface for searching and browsing over the resources provided by a large number of collections.

### 3.1 The NSDL Metadata Repository (MR)

The NSDL Metadata Repository (MR) is implemented as an Oracle™ relational database. Individual metadata records are stored in a series of tables, both as discrete Dublin Core metadata elements, and as a full XML metadata record. The database can support multiple metadata formats for a single record, and it maintains collection-item relationships, allowing identification of collection membership for all metadata records in the database.

Metadata records are aggregated from collection providers using the Open Archives Initiative Protocol for Metadata Harvesting (OAI-PMH) [18]. This protocol supports the incremental harvest of new and updated metadata records from the OAI provider. The ingest service of the MR regularly performs incremental harvests from all the collection providers, on a schedule commensurate with the frequency of updates for their collection.

During the ingest process, all metadata records are validated for conformance to the appropriate metadata schema. A group of *safe transforms* maps metadata elements to normalized, standard versions. These transforms also maps certain unqualified Dublin Core metadata elements to more detailed and explicit qualified Dublin Core elements. Finally, the metadata is transformed to the standard *nsdl_dc*[1] metadata format.

The MR also provides an OAI server[2], which allows other services to harvest the aggregated and normalized records from the MR. Currently, this is used by the search and archiving services to obtain the information needed both to build the search indexes used on the nsdl.org web site, and also to crawl and archive digital resources in the NSDL.

### 3.2 Services over the MR

Currently, there are two major services that make use of the MR. There is a search service, which uses the Lucene[3] full-text indexing system to index both the metadata records describing the resource and the text content of the first HTML page associated with the digital resource itself. There is also an archiving service, which does monthly web crawls of all the digital resources in the NSDL. The archiving service identifies a collection of linked pages which it believes best represents the actual resource, and it creates a snapshot archive of these pages.

Both services interact with the MR by using OAI-PMH to incrementally harvest new and updated metadata records from the MR. For the search service, these records are used both directly in building full-text indexes of the metadata and indirectly as a pointer to the digital content. The archiving service uses this information to update its local cache of URL starting points for its web crawl.

---

[1] http://ns.nsdl.org/schemas/nsdl_dc/nsdl_dc_v1.02.xsd

[2] http://services.nsdl.org:8080/nsdloai/OAI

[3] http://jakarta.apache.org/lucene/docs/index.html

Currently, these services provide access through REST (for the archiving service) and WEBDAV (for the search service) interfaces. The search service is currently being re-implemented, and one result of the new implementation will be a public and fully documented REST interface.

The fundamental model of service interaction for the current MR is through OAI-PMH harvesting. Services wishing to make use of the MR resources must harvest metadata records from it. Services that provide information to the MR must do so by creating an OAI server that the MR ingest process can harvest.

### 3.3 Experience with the MR

The MR has provided an excellent and robust initial framework for the NSDL. It currently harvests item level metadata records from over 75 OAI providers (collections), and it contains over 650,000 metadata records. The CI group has successfully implemented production-level searching, browsing, collection management, resource recommendation, and archiving services with the MR as the fundamental base.

As the CI group moves forward with new services, the MR limits this development in three ways: First, the underlying relational database and its table structure restrict the ability to directly support and manage digital content. It also creates a situation where one particular metadata format, Dublin Core, is treated as special by the MR: with DC elements broken out individually, while all other metadata types are just XML blobs. Second, there is no natural web services interface available to access and update records in the MR. This complicates and limits our ability to easily build new MR-based services. Third, the MR design only represents collection/item relationships. This limits our ability to represent contextual information for NSDL resources.

The MR was the right approach for the set of requirements that existed at the beginning of the project. However, with the new understanding of the importance of context for educational use and reuse of resources, those requirements have changed. To move forward, the NSDL needs a more flexible and expressive architecture.

### 4. A NETWORK OVERLAY

The notion of an *overlay network* has been applied in a number of contexts, most notably in the case of peer-to-peer (p2p) networks. Informally, an overlay network is a set of edges projected on top of a set of nodes that exist in some other network context – e.g., the Internet. Thus, a p2p overlay network is a set of edges, which represent the communication among peer nodes via some protocol (e.g., JXTA [38]) over the underlying Internet.

In the case of the NSDL, we are creating an overlay network that provides the information framework for adding value to STEM resources distributed over the Internet. Some requirements of this network are as follows:

- It must be resource-centric, rather than metadata-centric. The MR focused on representing metadata and relegated resources (content) to a second-class status. Intuitively a library *is* content and representing metadata is just one tool for managing that content. Management of those resources and their subsequent preservation, reuse, and persistent identification is only possible if they are fully represented in the underlying data model

- It must represent more than content resources, including with them other information that is part of the NSDL. This information includes the people, organizations, and services that contribute and use the resources; the standards that define grade and subject utility of the resources, curricula and lesson plans in which these resources have been used, and the like.

- It must combine information that is distributed over the Internet with information stored locally and within control of CI. In such a manner, the NSDL can act as a repository context for educators to reference, reuse, and add value to pre-existing primary resources.

- It must represent the diversity of relationships among the content, agents, services, standards, and the other entities in the NSDL context. The relationships must be expressed in an extensible ontology, so that other systems are able to use and reuse the relationship structure.

This section describes the design and implementation of a new NSDL architecture that stores and exposes an information network that meets these requirements. This architecture is implemented as a Fedora repository. For the purpose of backwards-compatibility with the MR architecture described in Section 3 this repository supports an OAI-PMH interface for harvesting metadata from OAI data providers and providing metadata to OAI service providers. The repository also supports a richer NSDL SOAP and REST interface for access to and addition of entities and relationships in the NSDL data model (e.g., resources, metadata, annotations, etc.) by services, organizations, and individuals that participate in the NSDL.

### 4.1 Fedora

Fedora is generally viewed as a "document storage system", in the manner of DSpace [32]. However, the rich object model underlying Fedora and exposure of the model through a web service interface makes Fedora usable for much more than simple document storage and delivery. This notion was introduced initially as value-added surrogates [25]. The remainder of this section summarizes the Fedora object model and provides the basis for understanding the NSDL content model implementation described in future sections. Readers interested on more details about Fedora should consult [17].

The Fedora digital object model allows the aggregation of local and distributed data in multiple formats. Web-accessible services may then be associated with the aggregated data. As a result a digital object is accessible in multiple representations, some of them direct transcriptions of aggregated data, and some of them produced dynamically by the associated web services.

The latest release of Fedora augments the digital object model by providing the infrastructure for expressing relationships among objects and their components. Examples of relationships between digital objects include well-known management relationships such as the organization of items in a collection, structural relationships such as the part-whole links between individual chapters and a book, and semantic relationships useful in educational digital library organization such as subject, grade, and curricula appropriateness. Fedora

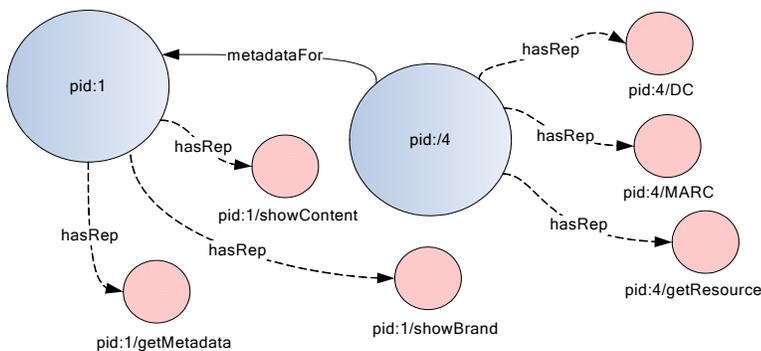

Figure 1 – Fedora representational view

All digital objects, and their individual representations, are identified with Uniform Resource Identifiers (URIs). These URIs are specified using the "info" scheme and conform to the syntax described at [2]. These URIs form the basis of Fedora's REST-based access service (i.e., API-A-LITE) that defines service request URLs for access to digital objects and their representations.

Figure 1 depicts the representational view of two inter-related Fedora objects. The figure shows a directed graph, where the larger blue nodes are digital objects (identified by *pids*), and the smaller pink nodes are representations of the digital objects. These nodes are linked by two types of arcs – (solid) relationship arcs connect digital objects, and (dashed) representation arcs connect digital objects to their respective representations. This graph can be expressed as RDF, stored in a triple store, and queried.

A brief summary of the figure, which is a simplified example of the NSDL data model that is described in Section 4.2, is as follows. The digital object labeled *pid:1* corresponds to an NSDL information resource with metadata represented by digital object *pid:4*. *pid:1* has three representations. The one labeled *pid:1/getMetadata* returns the identifier(s) of metadata objects associated with this resource. The one labeled *pid:1/showBrand* returns information about the branding (source) of this resource. The one labeled *pid:1/showContent* displays the content for this information resource. A relationship arc indicates that *pid:4* is metadata for *pid:1*. *pid:4* has representations, two of which are metadata in DC and MARC, and one that returns the resource that this metadata is about

We have yet to define the underlying source of these representations; i.e., where they are stored or whether they are stored statically or generated dynamically. In fact, in this view of the architecture such details are hidden from the client application concerned with access to these representations.

expresses relationships by defining a base relationship ontology using RDFS [8] and provides a slot in the digital object abstraction for RDF expression of relationships based on this ontology. Assertions from other ontologies may also be included along with the base Fedora relationships. All relationships are reflected in a native RDF triple-store using Kowari [36]. The query interface to this triple-store is exposed as a web service.

A Fedora repository is implemented as a set of web services and its full functionality, including its rich object model, is exposed through well-defined web service APIs. These APIs provides full programmatic management of information in the repository as well as search for and access to multiple representations of objects. As such, Fedora is particularly well-suited to co-exist in a broader web service framework and act as the foundation layer for a variety of application front-ends and user interfaces. In addition, Fedora comfortably co-exists in a heterogeneous environment by providing OAI-PMH access to the repository and supporting export and import of objects in multiple formats such as METS [3]and MPEG-DIDL [14]..

The Fedora software is the product of an ongoing research, implementation, and support project [33]. The roots lie in DARPA and NSF-funded research, with support for development of the open source implementation coming from the Andrew W. Mellon Foundation. Mellon-funded development continues through 2007.

The object model that lies at the core of Fedora can be understood from two perspectives; representational and functiona..

### 4.1.1 Representational perspective

The *representational* perspective defines a simplified graph-based abstraction of the Fedora object model. Each digital object is represented as a rooted sub-graph, modeling the notion that a digital object, the root of the sub-graph, can disseminate multiple *representations,* the other nodes in the sub-graph. The notion of a representation corresponds to the general notion of a complex object in METS [3] or DIDL [14] that has multiple views. A familiar example is an image available in multiple formats – e.g., gif, jpeg, tiff. Another example, employed in the NSDL data model, is an OAI metadata item, which can disseminate metadata in multiple formats (e.g., Dublin Core, MARC). These roots of these sub-graphs can then be connected via typed links, representing the relationships among digital objects.

### 4.1.2 Functional perspective

The functional perspective reveals the object components that underlie the representational perspective and provides the basis for understanding how the Fedora object model relates to the management services exposed in the Fedora repository architecture. A complete description of the functional view is out of scope for this paper. Interested readers should read Fedora-specific papers [17, 24, 26]. For the purpose of a brief summary, the functional components of the Fedora object model include:

- *Digital Objects* are uniquely-named information entities that are the foundation for aggregating data and associating services that process that data.

- *Datastreams* are information components of digital objects. A datastream may be *local,* with data contained within the digital object. Alternatively, a datastream may be *remote*, referencing external data via a URL. The externally referenced data may be a representation of another digital object. This provides the foundation for recursive reuse of primary and secondary resources and the composition of learning objects.

- *Disseminators* are sets of operations (methods), and metadata that establishes linkages among those operations, the web services that implement the operations, and the data in the digital object that provide input for the services. The linkage of a disseminator with a digital object endows it with the behavior defined by the disseminator, resulting in *virtual representations* – the output of the services that process the data. From the access perspective, these virtual representations of the digital object are indistinguishable from simple representations. Thus a digital object with a representation corresponding to some metadata format (e.g., Dublin Core), may disseminate that representation either by storing the DC metadata statically or by computationally deriving it from some other metadata format.
- *Relationship data* expresses relationships among Fedora digital objects as RDF fragments within a distinguished datastream in the digital object. These RDF fragments are merged into a joined graph of relationships and are represented in a Kowari triple-store [36]. This triple store can be queried directly via RDQL [31] through a web service interface.

The Fedora implementation wraps these functional components within a service-based architecture that supports versioning, integrity, access, and management of the digital objects.

## 4.2 Content Model for the Network Overlay

The notion of a *content model* has been used in the Fedora context [11] to describe the structuring of objects for a specific application. This object structure includes the configuration of datastreams, disseminators, and the relationship ontology. The expression of such a model is important for design purposes. In addition, if expressed formally, this model can be used to establish and enforce constraints on the data that instantiates the model.

In our NSDL work with Fedora we have used OWL [11] and Protégé [10] to formally express the content model for an information network overlay. The full details of this work are beyond the scope of this paper and will be described in future papers. This section provides a short summary.

One aspect of the content model is an object type hierarchy. The underlying Fedora architecture does not impose a type structure on digital objects. Instead, the disseminator component is the mechanism to associate operational semantics or types with digital objects. Since multiple disseminators, or sets of operational characteristics, may be linked to a digital object, Fedora digital objects are inherently polymorphic. For example, assuming a set of operations associate with *Metadata* – e.g., *getRecord*, *getResource* – and *Content* – e.g., *displayContent*, *getMetadata* – it is possible for a single digital object to have both types of operational behaviors. Both disseminators can be linked to a digital object, allowing it to "act like" both content and metadata.

The NSDL content model defines the following types and their primary operations, which should be understood within the context of this polymorphic typing. The next section demonstrates the use of these types in the NSDL information network.

- *Metadata* – with operations that return a record in a specified format (e.g., Dublin Core), return the provider of the metadata, and return the resource to which the metadata applies.
- *Resource* – with operations that return the handle identity of the resource (all NSDL resources are identified via the Handle System [1]), return the metadata objects that describe this resource, and sets into which this resource is aggregated. There are two sub-types of resource:
    - *Agent* – a person or organization associated with roles (see below) that the agent performs
    - *Content* – a resource that is an information unit in the NSDL with operations that allow access to the information.
- *Role* – an abstract super-class for sub-classes that define actions or services provided by agents. The notable operation for a role returns the brand for that role, which can then be associated with information objects linked with the role (e.g., provided metadata). There are two defined sub-classes for role:
    - *Aggregator* – indicating that the agent with this role forms sets of resources for reasons such as collection management, semantic grouping (standards, taxonomy), etc. This class has two core operations, one that returns resources that are members of the aggregation, and another that returns metadata about the aggregation.
    - *Metadata Provider* – indicating that the agent with this role provides metadata for resources (e.g., via OAI-PMH). This class has two core operations, one that returns the metadata provided (e.g., by OAI-PMH) by this agent, and another that returns a description of the method of provision.

Another component of the data model defines the ontology for expressing relationships among digital objects. The relationships in the NSDL content model are:

- *annotates* – relating a content resource to another resource. This indicates that the source content can be interpreted as a comment on the target resource.
- *assertedBy* – relating an agent and a role. This states that the target role is claimed by the source agent.
- *augments* – relating metadata and metadata. This indicates that the source metadata corrects and/or enhances the target metadata.
- *hasRole* – relating an agent and a role. This indicates that the source agent assumes the target role.
- *metadataFor* – relating metadata and a resource. This indicates that the source metadata is about the target resource.

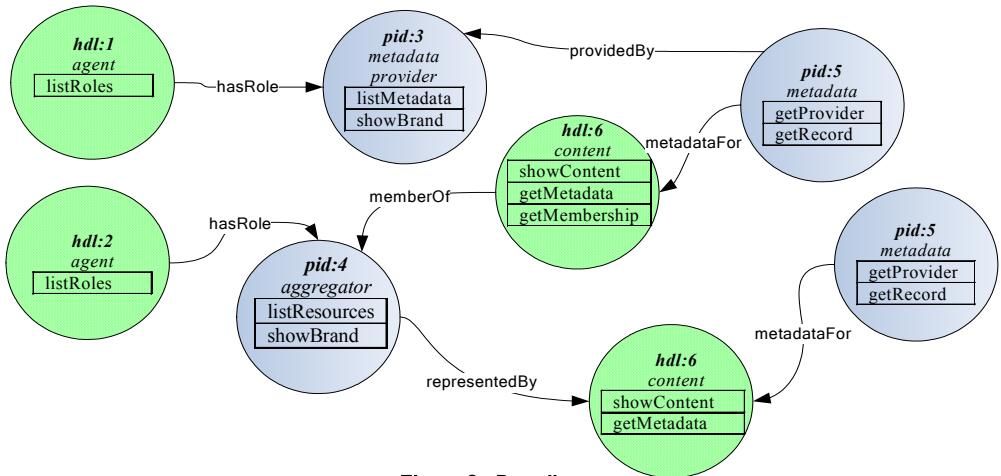

**Figure 2 - Branding**

## 4.3 Use cases for the content model

The classes and relationships described in the previous section provide the building blocks for modeling different NSDL use cases. This section describes a number of these use cases and the sub-graphs of the information network that model those use cases. For descriptive purposes, details of the actual graph are sometimes excluded from the examples.

The illustrated graphs have a number of shared characteristics that deserve explanation:

- The nodes correspond to digital objects in the Fedora repository.
- Nodes that are green have identifiers, or *hdls*, that are registered in the handle system [1], and thus are resources that have externally visible identities. Examples are agents, and content resources. Externally visible stable identifiers provide a basis for citation and other forms of external linking
- Nodes that are blue are internal to the model and only have Fedora identifiers or *pids*.
- Each node (digital object) contains text that is, in order, its identifier, type, and a table of its operations. This is a condensed version of the representational graph shown in Figure 1 – representation nodes are not shown. Instead

- *memberOf* – relating a resource and an aggregator. This indicates that the source resource is a member of the set defined by the target aggregator.
- *providedBy* – relating metadata and a metadata provider role. This indicates that the source metadata is provided by the target metadata provider.
- *representedBy* – relating an aggregator and a content resource. This indicates that the target resource is an information surrogate for the source aggregation. This is useful for defining a distinguished resource as a representation of a collection or aggregation. An example might be a content resource containing a state standard, that is linked to an aggregator that collects other NSDL resources relevant to this standard.

As described earlier, these inter-object relationships are expressed within a datastream of the digital objects and are joined in a searchable triple-store. Queries on the triple-store can be issued within a disseminator, and thereby the structure of the full relationship graph can be exposed as part of the operational semantics of objects.

The utility of relationship queries within disseminators is as follows. Note that the *memberOf* relationship between a resource and aggregator implies a reverse direction *hasMember* relationship that is not defined within the relationship ontology, and therefore is not stated in the relationship datastream of digital objects. Representing this relationship and similar relationships presents scaling problems due to bloating of the relationship datastream (i.e., an aggregation could have millions of members). Instead the semantics of the inverse relationship can be extracted via a query on the triple store using the *memberOf* relationship. This query can be included in a disseminator that exposes a *listMembers* operation for an aggregation.

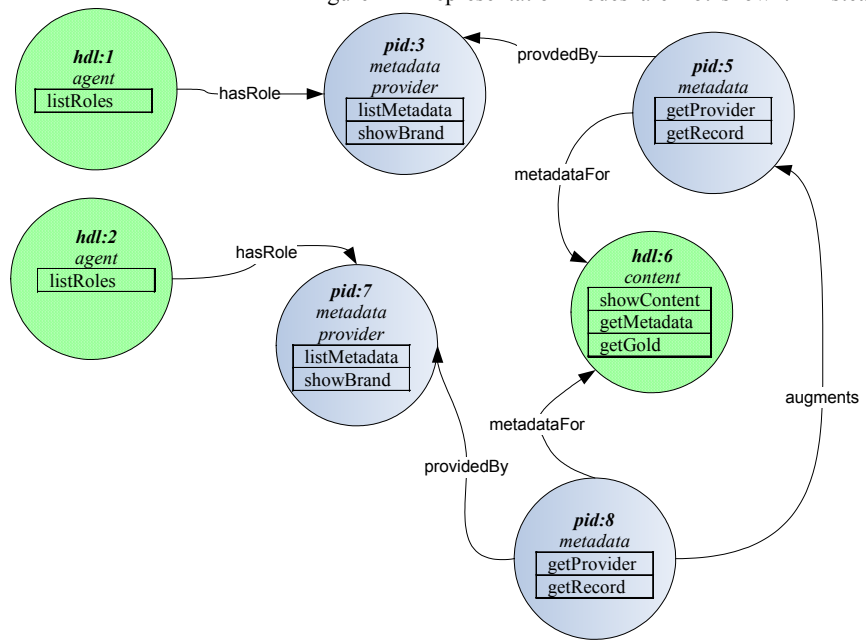

**Figure 3 – Multiple metadata sources**

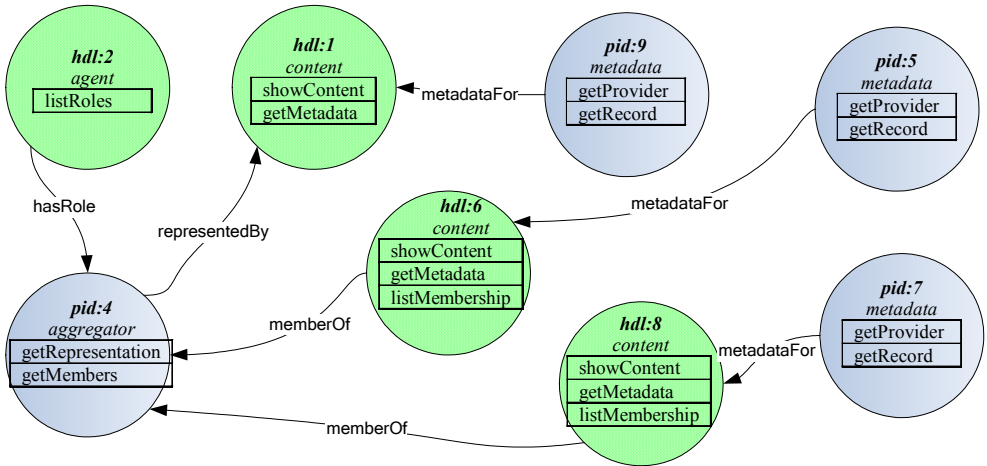

Figure 4 - Semantic aggregations

- each table entry corresponds to a node at the tail of a *hasRep* arc in that earlier graph. The list of operations corresponds to the operational semantics for the type of the object. For example, an "metadata" digital object has methods *getProvider* and *getRecprd*. To simplify the examples, only selected operations shown.
- The arcs between nodes are labeled with the relationship type.

### 4.3.1 Branding of resources and metadata

The agent, individual or organization that provided a resource and the reputation of that agent are important indicators of the quality of that information. Furthermore, the branding of metadata and the content that it describes are frequently distinct – certainly the source of library cataloging records is different from the books that they describe.

The graph shown in Figure 2 shows the sub-graph structure in the content model for associating information – metadata and content – with agents and their roles. The figure shows two agents, one a metadata provider, linked to a metadata object, and another an aggregator (resource selector), linked to a content object. The brands associated with each role provide branding information for each information resource. The brand for the metadata is that of the metadata provider. The brand for the resource is that of the aggregator to which the resource is linked as a member.

### 4.3.2 Multiple descriptions of resources

Improving metadata quality is a key issue for the NSDL and, in fact, for all digital libraries that harvest metadata from non-authoritative sources [13]. Among the techniques for improving metadata quality is associating metadata from multiple providers with resources and making it possible for one metadata provider to augment (correct, enhance, etc.) the metadata form another provider [12]. Layered services can then disseminate a computed "gold metadata record" that combines information from multiple sources based on their reputations.

The graph in Figure 3 shows the sub-graph structure in the content model for representing multiple metadata sourcing and augmentation. As indicated, there are two agents, each of which plays the role of metadata provider. Note that the content resource, *hdl:6*, has two metadata items, with the metadata in *pid:8* augmenting that in *pid:5*, and that the content resource has an operation *getGold*, which disseminates a computed "high quality" record.

### 4.3.3 Creating semantic aggregations

An essential means of organizing a library is grouping resources into various aggregations. These aggregations may define management units, such as collections, or semantic groupings, such as the association of a set of resources with a specific state standard or classification of them according to some subject taxonomy. As noted by Sumner, et. al. [34], the association of resources "can help educators and learners to locate, comprehend and use educational resources in digital libraries". The aggregation model needs to support the notion of a resource co-existing in multiple groupings – for example, a resource may apply to multiple state standards or it may be associated with multiple subject categories. It also needs to support recursive grouping to permit nested hierarchies.

Figure 4 illustrates the representation of an aggregation. As shown the agent represented by *hdl:2* is acting as an aggregator. The content resources *hdl:6* and *hdl:8* are members of the aggregation. The resource *hdl:1* is a content document representing the aggregation, such as a collection record or a standards document. The resources that are members of this aggregation may be in any number of other aggregations and, because the aggregation itself is represented by a resource, it may be grouped in other aggregations.

### 4.3.4 Annotations and reviews

Successful commercial sites such as Amazon.com demonstrate the utility of user reviews. When associated with reviewers and their respective reputations, such reviews can provide a useful foundation for sharing user experiences. In an educational context, the experiences of a fellow teacher with a learning object are frequently the most valuable information available about that resource.

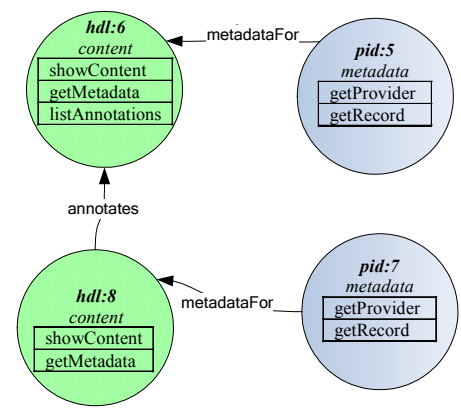

Figure 5 - Annotations and reviews

Figure 5 illustrates the representation of two resources, one that is a review of another. Note that the review is not represented as metadata because 1) reviews are not necessarily structured data in the form of metadata records and 2) the review itself is a content resource.

### 4.4 Initial Implementation Plans
As stated earlier, the first phase of the work described in this paper is due for release in the third quarter of 2005. The main goal of this release is to duplicate the functionality of the existing metadata repository (MR). Existing data providers that supply metadata to the NSDL via OAI-PMH and service providers that harvest metadata from the MR will be able to continue their work with no changes. From the user perspective, the operation of the NSDL portal at nsdl.org will remain unchanged.

The initial release will include three new pieces of functionality. First, CI will be able to ingest, store, and expose metadata in formats other than Dublin Core. Second, the new repository will offer metadata harvesters a new resource-centric aggregation metadata format, which will allow harvesters to access all the metadata that CI has about a specified resource. Finally, the implementation will provide the hooks for the integration of developing NSDL services that provide standard web services interfaces.

## 5. CONCLUSION AND FUTURE PLANS
The NSDL is intended to provide the context for improving education. As such, it needs to provide more than search and access to resources, which are only the basic functions of a digital library. It must, in addition, provide the framework for adding context to educational resources and support the reuse of those resources and their secondary products.

This paper has described an architecture and implementation by NSDL Core Integration that enables contextualization and reuse. This architecture is based on the notion of an information network overlay, which supports various entities with static and dynamic representations and embeds those entities in an ontology-based relationship graph. The network overlay is implemented by a Fedora repository and exposed as a set of web services. While the architecture and data model is designed for the NSDL, we believe that the notion of an information network overlay extends to other digital libraries that aggregate distributed resources and add value to them.

Phase I implementation plans call for duplication of existing metadata repository functionality with some additions to that functionality. Future releases of the new NSDL repository will expose the richer features expressed the data model described in this paper. These include standards alignment, subject classification, reviews, and reputation systems. OAI-PMH, the single mode of access to the current MR, is not sufficiently expressive for access to this richer inter-related information. As such, future releases will expose two new interfaces to the repository:

- A SOAP [7] and/or REST-based API that exposes operations on the entities in the NSDL content model – e.g., agents, aggregations, content, etc.
- A SOAP and/or REST-based RDQL [31] query interface to the repository relationship graph.

NSDL portal builders, such as the Pathways projects, will be able to use these interfaces for the development of rich user-visible services that build on the infrastructure provided by CI.

The success of this architecture has admittedly yet to be proven. However, we are confident that the data model described in this paper and its Fedora implementation will act as the catalyst for a richer and more educationally useful NSDL.

## 6. ACKNOWLEDGMENTS

The authors acknowledge the support of the entire NSDL community, notably Lee Zia and Kaye Howe. This work is funded by the National Science Foundation under grant numbers 0127308 and 0127520. Fedora work is funded by a grant from the Andrew W. Mellon Foundation. The authors acknowledge the contributions of the Fedora project team, especially Sandy Payette and Chris Wilper at Cornell.